\documentclass[a4paper]{jpconf}
\usepackage{graphicx}
\begin{document}
\title{Unified Description of Classical and Quantum Behaviours in a Variational Principle}

\author{Tomoi Koide,    Takeshi Kodama}
\address{Instituto de F\'{\i}sica, Universidade Federal do Rio de Janeiro, C.P. 68528,
21941-972, Rio de Janeiro, Brazil}
\ead{tomoikoide@gmail.com, kodama.takeshi@gmail.com}
\author{Kazuo Tsushima}
\address{International Institute of Physics, Federal University of Rio Grande do Norte,
Natal 59078-400, RN, Brazil}
\ead{kazuo.tsushima@gmail.com}

\begin{abstract}
We give a pedagogical introduction of the stochastic variational method and
show that this generalized variational principle describes classical and quantum mechanics
in a unified way.
\end{abstract}

\section{Introduction} \

Variational approach conceptually plays a fundamental role in elucidating
the structure of classical mechanics, clarifying the origin of dynamics and
the relation between symmetries and conservation laws.
In classical mechanics, the optimized function is characterized by Lagrangian,
defined as $T-V$ with $T$ and $V$ being a kinetic and a potential terms, respectively.

We can still argue the variational principle in quantum mechanics, but the
Lagrangian does not have any more the form of $T-V$, instead it is given by
$\psi^* (i\hbar \partial_t - \hat{H}) \psi$, where $\hat{H}$ is a Hamiltonian operator
and $\psi$ is a wave function.
Therefore, at first glance, any clear or direct correspondence between classical and quantum mechanics
does not seem to exist in the variational point of view, but it does exist.
If we extend the idea of the variation to stochastic variable, the variational principle describes classical and quantum behaviors in a unified way.

This method is called stochastic variational method (SVM) and firstly proposed by Yasue \cite{yasue,guerra1,guerra2,guerra3,guerra4} so as to reformulate
Nelson's stochastic quantization~\cite{nelson1,nelson2}.
This framework is, however, based on special techniques attributed to stochastic calculus which is not familiar to physicists.
In this paper, we give a pedagogical introduction of SVM in a self-contained manner,
showing the unified description of classical and quantum mechanics.
As another review, see, for example, Ref. \cite{zam}.

\newpage

\section{Variational method for stochastic variables} \

Because of the limitation of pages, we cannot explain all aspects of stochastic calculus in detail.
See for example, Ref.~\cite{gardinar} for standard techniques which are not explained here.

\subsection{Forward and Backward SDEs} \

In the variational principle for stochastic variables, 
a particle trajectory is not any more smooth and given by a zig-zag path in general.
As the consequence, the evolution of a particle trajectory is defined by the following stochastic differential equation (SDE),
\begin{equation}
d{\bf r}(t) = {\bf u}({\bf r}(t),t)dt + \sqrt{2\nu}d{\bf W}_t~~~~(dt > 0). \label{fsde}
\end{equation}
In this paper, a difference $d A(t)$ is always defined by $A(t+dt) - A(t)$ independently of the sign of $dt$.
The last term in Eq.~(\ref{fsde}) is the origin of the zig-zag motion and called noise term. The parameter $\nu$
characterizes the strength of this noise term. One can easily see that ${\bf u}({\bf r}(t),t)$ is reduced to
the usual classical definition of the particle velocity in the limit of vanishing $\nu$.
The property of ${\bf W}_t$ depends on the stochastic property of the noise term.
In the present paper, we assume that ${\bf W}_t$ is the Wiener process, which is characterized
by the following correlation properties,
\begin{eqnarray}
E[d{\bf W}_t] &=& 0, \\
E[(dW^i_t)(dW^j_t)] &=& |dt| \delta^{ij},~~(i,j = x,y,z), \label{corr2}\\
E[W^i_t dW^j_{t'}] &=& 0~~{\rm for}~~(t\le t'),
\end{eqnarray}
where $E[~~~]$ indicates the average of stochastic events.
It is clear from the above properties that $d{\bf W}_t$ behaves as the so-called Gaussian white noise.

Such a SDE (Langevin equation) has been used in statistical physics to discuss, for example, thermalization.
This is essentially an irreversible process and we exclusively discuss the time evolution for a given initial condition.
However, in the formulation of a variational method, we should fix not only an initial condition but also a final condition.
If we consider a backward process in time, $dt <0$, it should describe a stochastic process from
the final condition to the initial condition.

Then what is the time-reversed process corresponding to Eq.~(\ref{fsde})? To discuss this, let us define the
probability distribution as
\begin{equation}
\rho({\bf x},t) = \int d^3 {\bf r}_i~\rho_I ({\bf r}_i) E[ \delta^{(3)}({\bf x} - {\bf r}(t)) ],
\end{equation}
where ${\bf r}(t)$ is the solution of Eq.~(\ref{fsde}) and $\rho_I ({\bf r}_i)$ is the initial particle distribution with
${\bf r}(t_i) = {\bf r}_i$ at an initial time $t_i$. As is well-known, the evolution equation is
given by the Fokker-Planck equation,
\begin{equation}
\partial_t \rho ({\bf x},t) = \nabla (-{\bf u} ({\bf x},t) + \nu \nabla) \rho ({\bf x},t). \label{ffp}
\end{equation}

If the probability distribution evolves from $\rho_I ({\bf r})$ to $\rho_F ({\bf r}) \equiv \rho({\bf r}(t_f),t_f)$ at a final time $t_f$ following
Eq.~(\ref{ffp}), the corresponding time-reversed process should describe the evolution from $\rho_F$ to $\rho_I$.
Suppose that this process is described by
\begin{equation}
d{\bf r}(t) = \tilde{\bf u}({\bf r}(t),t)dt + \sqrt{2\nu}d{\bf W}_t, \label{bsde}
\end{equation}
where, it should be noted, $dt < 0$. Then differently from the classical dynamics, it generally holds,
\begin{equation}
\tilde{\bf u}({\bf r},t)dt \neq - {\bf u}({\bf r},t)|dt|. \label{cond-uso}
\end{equation}
To understand this reason, let us consider the case ${\bf u}=0$, where
Eq.~(\ref{ffp}) becomes a simple diffusion equation, and hence the corresponding time-reversed process
should describe an accumulation (opposite of diffusion) process.
However, if the condition~(\ref{cond-uso}) were satisfied,
Eq.~(\ref{bsde}) coincides with the diffusion equation and cannot describe the accumulation process.

To obtain the precise relation instead of Eq.~(\ref{cond-uso}), we calculate the Fokker-Planck equation
assuming Eq.~(\ref{bsde}) as
\begin{equation}
\partial_t \rho ({\bf x},t) = \nabla (-\tilde{\bf u} ({\bf x},t) - \nu \nabla) \rho ({\bf x},t). \label{bfp}
\end{equation}
For the two Fokker-Planck equations (\ref{ffp}) and (\ref{bfp}) to be equivalent,
we find the following condition
\begin{equation}
{\bf u}({\bf x},t) = \tilde{\bf u}({\bf x},t) + 2\nu \nabla \ln \rho. \label{cc}
\end{equation}
This is the consistency condition for Eq.~(\ref{bsde}) to be the time-reversed process of Eq.~(\ref{fsde}).
\footnote{Strictly speaking, there is an ambiguity to obtain this relation. See Ref. \cite{kk2}.}
In fact, for the diffusion case where ${\bf u}=0$, we obtain by the consistency condition,
\begin{equation}
\tilde{\bf u}({\bf x},t) = - 2\nu \nabla \ln \rho.
\end{equation}
Substituting this, one can see that Eq.~(\ref{bsde}) indeed describes the accumulation process.
Interestingly, this consistency condition can be derived even from the property of Bayes' theorem.
See Ref. \cite{cati}.

In the following, we call Eq.~(\ref{fsde}) forward stochastic differential equation (SDE),
and Eq.~(\ref{bsde}) backward SDE, respectively.

For the later convenience,
let us introduce the mean velocity,
\begin{equation}
{\bf v}({\bf x},t) = \frac{{\bf u}({\bf x},t) + \tilde{\bf u}({\bf x},t)}{2}. \label{vm}
\end{equation}
Once this quantity is determined, one can easily find ${\bf u}$ and $\tilde{\bf u}$ by using the consistency condition.
This mean velocity is parallel to the flow of the particle probability distribution.
In fact, the above two Fokker-Planck equations are reduced to the following simple equation,
\begin{equation}
\partial_t \rho ({\bf x},t) = - \nabla \cdot (\rho ({\bf x},t) {\bf v}({\bf x},t)). \label{fp-true}
\end{equation}

\subsection{Velocities in SVM and partial integration formula} \

The action which we will optimize is the time integral of the Lagrangian, which
depends on the particle velocity. When the trajectory is described
by stochastic variables, however, the definition corresponding to the velocity is not trivial.

As is well-known, the time derivative of the trajectory described by SDE, for example Eq.~(\ref{fsde}),
is not well-defined in the limit of $|dt| \rightarrow 0$. This can be seen from the fact that $d{\bf W}_t$
has a size proportional to $\sqrt{|dt|}$ from Eq.~(\ref{corr2}), and thus $d{\bf r}/dt \sim d{\bf W}_t /dt \sim 1/\sqrt{|dt|}$.

However, it is known that there are two possible definitions which have a well-defined limit of $dt$ proposed by Nelson~\cite{nelson1,nelson2}:
One is the the mean forward derivative
\begin{equation}
D {\bf r}(t) = \lim_{dt \rightarrow 0+} E \left[ \frac{{\bf r}(t + dt) - {\bf r}(t)}{dt} \Big| {\cal P}_t \right],
\end{equation}
and the other the mean backward derivative,
\begin{equation}
\tilde{D} {\bf r}(t) = \lim_{dt \rightarrow 0-} E \left[ \frac{{\bf r}(t + dt) - {\bf r}(t)}{dt} \Big| {\cal F}_t \right].
\end{equation}
These expectations are conditional averages, where ${\cal P}_t$ (${\cal F}_t$) indicates to fix values of ${\bf r}(t')$ for $t' \le t~~(t' \ge t)$.
For the Wiener process, one can easily find $D {\bf W}_t = 0$, but, in general, $\tilde{D}{\bf W}_t \neq 0$.\footnote{To understand this, the argument around Eq.~(\ref{example-ito}) will be useful.}
When ${\bf r}(t)$ is described by the forward and backward SDEs defined above,
we find $D {\bf r}(t) = {\bf u}({\bf r}(t),t)$ and
${\tilde D} {\bf r}(t) = \tilde{\bf u}({\bf r}(t),t)$.

As a matter of fact, it is impossible to control the behavior of each trajectory completely
because of the random noise. What we can adjust is, at best, only the trend of stochastic motions.
Then the mean forward derivative defined above represents the most probable velocity forward in time
when a particle is located at ${\bf r}(t)$, and the mean backward derivative is that of the backward in time.
Therefore, what we should obtain by the variational procedure is the form of ${\bf u}$
(or equivalently $\tilde{\bf u}$), and it is natural to express the velocities appearing in an optimized
function by these quantities.

Because of the two different time derivatives, the partial integration formula for the stochastic variable is
modified as
\begin{equation}
\int^t_0 ds E\left[ (DX(s)) Y(s) \right] = E\left[ X(t)Y(t) - X(0) Y(0)\right] - \int^t_0 ds E\left[ X(s) \tilde{D} Y(s)    \right].
\end{equation}
The derivation is given in \ref{pif}.
One should notice that when the time derivative $D$ moves from the left variable
($X(t)$) to the right ($Y(t)$), it is replaced by $\tilde{D}$.

\section{Variation of stochastic action} \

As an example of SVM, let us consider the optimization of the one particle Lagrangian,
\begin{equation}
L = \frac{m}{2}\dot{\bf r}^2(t)  - V({\bf r}(t)), \label{cla-lag}
\end{equation}
where $m$ is the mass of the particle and $V$ is a potential.
As is well-known, Newton's equations of motion is obtained when the usual variational method is applied to this.
To implement the stochastic variation to this Lagrangian, we need to express each term by
the corresponding stochastic quantities.

Due to the existence of the two possible definitions of the time derivatives,
the most general quadratic form of the kinetic energy of the Lagrangian is given by
\cite{koide}
\begin{equation}
\frac{m}{2} \dot{\bf r}^2(t) \longrightarrow
\frac{m}{2} \left[B_+ \{ A_+ (D{\bf r}(t))^2 + A_- (\tilde{D}{\bf r}(t))^2 \}
+ B_- (D{\bf r}(t))\cdot (\tilde{D}{\bf r}(t))
\right],
\end{equation}
where $A_\pm = 1/2 \pm \alpha_1$ and $B_\pm = 1/2 \pm \alpha_2$ with $\alpha_1$ and $\alpha_2$ being arbitrary real constants.
Note that the right hand side reduces to the left hand side in the limit $\nu \to 0$,
independently of the values of $\alpha_i~(i=1,2)$.
If $\alpha_1 \neq 0$, the optimized dynamics violates the time reversal symmetry and we obtain, for example, the Navier-Stokes-Fourier equation~\cite{kk2}.
In the present discussion, however, we focus on dynamics with the time-reversal symmetry, and choose $(\alpha_1,\alpha_2) = (0,1/2)$.
Then, the stochastic action corresponding to Eq.~(\ref{cla-lag}) is given by
\begin{equation}
I[{\bf r}(t)]
= \int^{t_f}_{t_i} dt E \left[ L({\bf r}, D{\bf r}, \tilde{D}{\bf r}) \right]=
\int^{t_f}_{t_i} dt E \left[ \frac{m}{4}((D{\bf r}(t))^2 + (\tilde{D}{\bf r}(t))^2) - V({\bf r}(t))\right]. \label{sto-action}
\end{equation}
It is known that there are several definitions for products of stochastic variables, for example, the Ito definition, Stratonovich definition and so on. 
However, there is no this ambiguity for, for example, $(D{\bf r}(t))^2$ in this formulation. 
See \ref{app:ito-st} for details.

The variation of the stochastic variable is introduced as
\begin{equation}
{\bf r}(t) \longrightarrow {\bf r}(t) + {\bf f}({\bf r}(t),t).
\end{equation}
Here ${\bf f}({\bf x},t)$ is an arbitrary infinitesimal function satisfying ${\bf f}({\bf x},t_i) = {\bf f}({\bf x},t_f) = 0$.
Then, for example, the variation of the kinetic term is calculated as
\begin{eqnarray}
\lefteqn{\int^{t_f}_{t_i} dt E \left[ (D{\bf r}(t) + D{\bf f}({\bf r}(t),t))^2 \right] - \int^{t_f}_{t_i} dt E \left[ (D{\bf r}(t))^2 \right] } && \nonumber \\
&& = 2 \int^{t_f}_{t_i} dt E \left[ {\bf u}({\bf r}(t),t)\cdot (D{\bf f}({\bf r}(t),t)) \right]
+ O({\bf f}^2) \nonumber \\
&& = - 2 \int^{t_f}_{t_i} dt E \left[ \{\tilde{D}{\bf u}({\bf r}(t),t)\} \cdot {\bf f}({\bf r}(t),t) \right].
\end{eqnarray}
Here we have first used the definition of the mean forward derivative, and then
the stochastic partial integration formula.
The potential part does not contain any time derivative terms, and its variation is the same as that
in the classical variational method.
Then the result of the variation is obtained as
\begin{eqnarray}
\delta I = \int^{t_f}_{t_i} dt E\left[
\left\{
-\frac{m}{2}( \tilde{D}{\bf u}({\bf r}(t),t) + D \tilde{\bf u}({\bf r}(t),t) ) - \nabla V({\bf r}(t))
\right\}\cdot {\bf f}({\bf r}(t),t)
\right].
\end{eqnarray}
It is clear from the definition of the mean derivatives that ${\bf r}(t)$ in $\tilde{D}{\bf u}$ is described by
the {\it backward} SDE. Then, substituting the definition of $\tilde{D}$ and applying Ito's lemma (\ref{app:ito}), we obtain
\begin{equation}
\tilde{D}{\bf u}({\bf r}(t),t) =  \left( \partial_t  + \tilde{\bf u}({\bf r}(t),t)\cdot \nabla - \nu \nabla^2 \right){\bf u}({\bf r}(t),t) . \label{d-cal1}
\end{equation}
Similarly, ${\bf r}(t)$ in $D\tilde{\bf u}$ is given by the {\it forward} SDE leading to
\begin{equation}
 {D}\tilde{\bf u}({\bf r}(t),t) = \left( \partial_t  + {\bf u}({\bf r}(t),t)\cdot \nabla + \nu \nabla^2 \right)\tilde{\bf u}({\bf r}(t),t) .\label{d-cal2}
\end{equation}
Note that the last noise term in Ito's lemma disappears in the above expressions,
because of the conditional average included in the definition of the mean derivatives.
In fact,
\begin{equation}
E \left[ \sqrt{2\nu}d{\bf W}_t \cdot \nabla  {\bf u}({\bf r}(t),t)  \Big| {\cal F}_t \right]
= \sqrt{2\nu} \nabla  {\bf u}({\bf r}(t),t) \cdot  E \left[d{\bf W}_t  \right] = 0~~(dt < 0). \label{example-ito}
\end{equation}

In the variational principle of stochastic variables,
we require that $\delta I$ vanishes for 1) any choice of ${\bf f}({\bf x},t)$, and also 2) any distribution of the stochastic variable ${\bf r}(t)$.
To satisfy these, ${\bf u}$ (or equivalently $\tilde{\bf u}$) should be the solution of
\begin{equation}
\left[
-\frac{m}{2}( \tilde{D}{\bf u}({\bf r}(t),t) + D \tilde{\bf u}({\bf r}(t),t) ) - \nabla V({\bf r}(t))
\right]_{{\bf r}(t) = {\bf x}} = 0. \label{sel2}
\end{equation}
Substituting Eqs. (\ref{d-cal1}) and (\ref{d-cal2}), we obtain
\begin{equation}
(\partial_t + {\bf v} ({\bf x},t)\cdot \nabla){\bf v} ({\bf x},t) = - \frac{1}{m}\nabla V({\bf x}) + 2\nu^2 \nabla \rho^{-1/2}({\bf x},t) \nabla^2 \sqrt{\rho({\bf x},t)} , \label{qhydro}
\end{equation}
where the mean velocity ${\bf v}$ is defined by Eq. (\ref{vm}).

It is worth mentioning that Eq. (\ref{sel2}) is formally expressed as
\begin{eqnarray}
\left[ \tilde{D} \frac{\partial L}{\partial D {\bf r}(t)} + D \frac{\partial L}{\partial \tilde{D} {\bf r}(t)}
- \frac{\partial L}{\partial {\bf r}(t)} \right]_{{\bf r}(t) = {\bf x}} = 0 .\label{sel}
\end{eqnarray}
Note that the stochastic variable ${\bf r}(t)$ is replaced by the position parameter ${\bf x}$
in the above, only after operating all mean derivatives.
This is nothing but the stochastic generalization of the Euler-Lagrange equation.

\section{Schr\"{o}dinger equation} \

These two equations (\ref{fp-true}) and (\ref{qhydro}) determine the optimized dynamics of the action given by Eq.~(\ref{sto-action}).
However, these coupled equations can be cast into a more familiar form.
Let us introduce the following complex function,
\begin{equation}
\psi ({\bf x},t) = \sqrt{\rho ({\bf x},t)}e^{i\theta ({\bf x},t)},
\end{equation}
where the phase is defined by
\begin{equation}
{\bf v} ({\bf x},t)= 2\nu \nabla \theta  ({\bf x},t).
\end{equation}
Then, from Eqs. (\ref{fp-true}) and (\ref{qhydro}), the evolution equation of this quantity is given by
\begin{equation}
i\partial_t \psi ({\bf x},t) = \left[ -\nu \nabla^2 + \frac{1}{2\nu m}V ({\bf x}) \right] \psi ({\bf x},t).
\end{equation}
When we choose $\nu = \hbar/(2m)$, this is reduced to the Schr\"{o}dinger equation, and
$\psi ({\bf x},t)$ is identified with the wave function.
Furthermore, one can easily find that $|\psi({\bf x},t)|^2$ gives the probability
density distribution, without introducing any quantum mechanical interpretations.
In short, the procedure described above, can be regarded to give an alternative quantization scheme.

\section{Stochastic Noether theorem} \

In the SVM quantization scheme,
the physical operators are defined through Noether's theorem for the stochastic action~\cite{misawa3}.
Let us consider the spatial translation by an arbitrary time-independent spatial vector ${\bf A}$ as 
${\bf r}(t) \longrightarrow {\bf r}(t) + {\bf A}$.
Now let us introduce the difference of the stochastic actions\ (\ref{sto-action}) before and after the transform by 
$\delta I =  I[{\bf r}(t) +{\bf A}] - I[{\bf r}(t)]$.
For the infinitesimal transform of ${\bf A}$, this quantity is given by 
\begin{eqnarray}
\delta I 
&=& 
\int^{t_f}_{t_i} dt E\left[ L ({\bf r}(t)+{\bf A}, D{\bf r}(t),\tilde{D}{\bf r}(t)) \right]
-
\int^{t_f}_{t_i} dt E\left[ L ({\bf r}(t), D{\bf r}(t),\tilde{D}{\bf r}(t))\right] \nonumber \\
&=&
\int^{t_f}_{t_i} dt E\left[  \frac{\partial L}{\partial {\bf r}} \cdot {\bf A}
\right] + O ({\bf A}^2) 
=
\int^{t_f}_{t_i} dt E\left[ \tilde{D} \frac{\partial L}{\partial D {\bf r}} 
+ D \frac{\partial L}{\partial \tilde{D} {\bf r}} 
\right] \cdot {\bf A} 
\nonumber \\
&=& \frac{m}{2}\int^{t_f}_{t_i} dt \frac{d}{dt} E \left[  D{\bf r} + \tilde{D}{\bf r} \right] \cdot{\bf A}.
\end{eqnarray}
Here, in the second line, we used the stochastic Euler-Lagrange equation (\ref{sel}) and, 
from the second to the third line, we have 
\begin{equation}
\frac{d}{dt}E\left[ X Y \right] = E \left[ Y DX + X \tilde{D} Y \right],
\end{equation}
which is obtained from the stochastic partial integration formula.

Suppose that our action is invariant for any homogeneous spatial translation, $\delta I=0$. 
In such a case, one can deduce that the quantity,
$m E \left[  (D{\bf r} + \tilde{D}{\bf r})/2 \right] $, is conserved.
Using the solution obtained by the stochastic variation,
this quantity is expressed as
\begin{equation}
\int d^3 {\bf r}_i~\rho_I\ \frac{m}{2} E \left[  D{\bf r} + \tilde{D}{\bf r} \right]
= \int d^3 {\bf x} \rho ({\bf x},t) m {\bf v} ({\bf x},t)
= \int d^3 {\bf x} \psi({\bf x},t) (-i\hbar \nabla) \psi ({\bf x},t).
\end{equation}
Here the conserved quantity is integrated for the initial particle distribution.
This is the well-known expression of momentum expectation value in quantum mechanics
and $-i\hbar \nabla$ is identified with the momentum operator.
Similarly, we can obtain the conservation laws of energy, angular momentum and charge by the stochastic Noether theorem.

\section{Canonical equation} \

Although the canonical formulation of SVM has not yet been established, we can still formally write down the stochastic canonical equation~\cite{kk3}.
Let us introduce quantities corresponding to the momenta as
\begin{equation}
\frac{1}{2} {\bf p} = \frac{\partial L }{\partial D{\bf r}},~~~~~~
\frac{1}{2} \bar{\bf p} = \frac{\partial L }{\partial \tilde{D}{\bf r}}. \label{defp}
\end{equation}
Then the stochastic Hamiltonian can be introduced by the Legendre transform as
\begin{equation}
H ({\bf r},{\bf p}, \bar{\bf p})
= \frac{1}{2} ({\bf p}\cdot D{\bf r} + \bar{\bf p} \cdot \tilde{D}{\bf r} )
- L({\bf r}, D{\bf r}, \tilde{D}{\bf r}). \label{ham}
\end{equation}
Substituting the stochastic Lagrangian used in Eq. (\ref{sto-action}), we obtain 
$H ({\bf r},{\bf p}, \bar{\bf p})= ({\bf p}^2 + \bar{\bf p}^2)/(4m) + V({\bf r})$, but we do not need this explicit form in the following discussion.

The variables $D{\bf r}$ and $\tilde{D}{\bf r}$ are now functions of ${\bf r}$, ${\bf p}$ and $\bar{\bf p}$.
To find the relations, we consider the following transforms,
${\bf r}  \longrightarrow  {\bf r} + \mbox{\boldmath$\eta$},\
{\bf p}  \longrightarrow  {\bf p} + \mbox{\boldmath$\zeta$}, \
\bar{\bf p}  \longrightarrow  \bar{\bf p} + \bar{\mbox{\boldmath$\zeta$}}$,
where \mbox{\boldmath$\eta$}, \mbox{\boldmath$\zeta$} and $\bar{\mbox{\boldmath$\zeta$}}$
are infinitesimal constants.
The both sides in Eq.\ (\ref{ham}) are expressed up to the first order as
\begin{eqnarray}
{\rm L.\ H.\ S.} &=& H  + \frac{\partial H}{\partial {\bf r}} \cdot \mbox{\boldmath$\eta$} + 
\frac{\partial H}{\partial {\bf p}} \cdot \mbox{\boldmath$\zeta$} +
\frac{\partial H}{\partial \bar{\bf p}} \cdot \bar{\mbox{\boldmath$\zeta$}} ,\\
{\rm R.\ H.\ S.} &=& 
H - \frac{\partial L}{\partial {\bf r}} \cdot \mbox{\boldmath$\eta$} + \frac{1}{2} D{\bf r} \cdot \mbox{\boldmath$\zeta$}
+ \frac{1}{2}  \tilde{D}{\bf r} \cdot \bar{\mbox{\boldmath$\zeta$}} ,
\end{eqnarray}
respectively. For the calculation of R. H. S., we used Eq.\ (\ref{defp}), and 
the result that $D{\bf r}$ is transfomred as 
\begin{eqnarray}
D {\bf r} &\longrightarrow& D{\bf r} + \sum_j \left(\frac{\partial D{\bf r}}{\partial r_j}\eta_j 
+  \frac{\partial D {\bf r}}{\partial p_j}\zeta_j
+  \frac{\partial D {\bf r}}{\partial \bar{p}_j}\bar{\zeta_j}
\right).
\end{eqnarray}
Much the same is true on $\tilde{D} {\bf r}$.

Both sides of Eq.\ (\ref{ham}) should coincide and thus we find the following relations:
\begin{eqnarray}
\frac{\partial H({\bf r},{\bf p} , \bar{\bf p}) }{\partial {\bf r}}  = - \frac{\partial L}{\partial {\bf r}},~~
\frac{\partial H({\bf r},{\bf p} , \bar{\bf p}) }{\partial {\bf p}} = \frac{1}{2}D{\bf r},~~
\frac{\partial H({\bf r},{\bf p} , \bar{\bf p}) }{\partial \bar{\bf p}} = \frac{1}{2} \tilde{D} {\bf r}. \label{p2}
\end{eqnarray}
Using this first relation and the definitions of the momenta,
we can re-express the stochastic Euler-Lagrange equation as 
\begin{eqnarray}
\tilde{D} \frac{\partial L}{\partial D {\bf r}} + D \frac{\partial L}{\partial \tilde{D} {\bf r}} - \frac{\partial L}{\partial {\bf r}} = 0
\longrightarrow \frac{1}{2} \tilde{D} {\bf p} + \frac{1}{2} D \bar{\bf p}
+ \frac{\partial H({\bf r},{\bf p} , \bar{\bf p}) }{\partial {\bf r}} = 0 .
\end{eqnarray}
This and the last two of Eq.\ (\ref{p2}) correspond to the canonical equations. By substituting this into Eq.\ (\ref{sel})
with the help of the second and the last equations in Eq.\ (\ref{p2}),
one can confirm that Eq.\ (\ref{qhydro}) is indeed reproduced.

\section{Concluding remarks} \

We have discussed the application of SVM to quantize classical particle systems.
When we apply the stochastic variation of the action~(\ref{sto-action}) assuming Eqs.~(\ref{fsde})
and~(\ref{bsde}), we obtain the Schr\"{o}dinger equation.
The result of the variation depends on the assumed form of Eqs.~(\ref{fsde})
and~(\ref{bsde}). If we use them with the limit of $\nu_i \rightarrow 0$,
the stochastic variation of the same action leads to Newton's equation of motion.
That is, the usual variational method is a special case of SVM,
and both classical and quantum mechanics are described in the framework of this more generalized
variational method.
It is also possible to apply SVM to quantize Klein-Gordon field~\cite{koidekg1} and abelian gauge field~\cite{koidekg2}.
As a related work associated with quantized fields and random fields, see Ref. \cite{morgan}.

The framework of SVM itself can be regarded more general than the method of quantization.
In fact, it is possible to derive the Navier-Stokes-Fourier equation
by applying SVM to the action, which leads to the Euler equation when the usual classical mechanical
method of variation is applied~\cite{kk2}.
It is interesting to note that the Gross-Pitaevskii equation also can be obtained in the
framework of SVM \cite{kk2}.

There are various proposals for the non-conventional quantization scheme.
One of them is the so-called stochastic quantization proposed by Parisi and Wu \cite{parisi,namiki}.
In a similar way to SVM, the effect of quantum fluctuation is taken into account through SDE
even in this method, but the philosophy for quantization seems to be completely different.
For example, a fictitious time variable is introduced in the stochastic quantization.
That is, to quantize a $3+1$ dimensional system, we need to consider $3+1+1$ dimension.
Then SDE describes the evolution in this fictitious time.
Moreover, what is calculated in this approach is propagators while the Schr\"{o}dinger equation and
physical operators are obtained in SVM.
For other quantizations, See, for example, Refs. \cite{cati,otherquan1,otherquan2,otherquan3,otherquan4,otherquan5}.

Finally, we would like to list up future problems to be studied by SVM,

\begin{itemize}
  \item Quantization of fermions
  \item Criticism by Takabayashi \cite{taka1,taka2} and Wallstrom \cite{wall1,wall2}
  \item Canonical transform \cite{misawa-cano}
  \item Extension to general curved coordinate systems
  \item Classicalization and quantum-classical hybrids \cite{koide-qch}
  \item Variational formulation of relativistic dissipative fluids
  \item Topology
  \item Anomaly
\end{itemize}

\ack

This work is financially supported by CNPq.
KT is supported by the Brazilian Ministry of Science,
Technology and Innovation (MCTI-Brazil), and
Conselho Nacional de Desenvolvimento Cient{\'i}fico e Tecnol\'ogico
(CNPq), project 550026/2011-8.

\appendix

\section{Stochastic partial integration formula} \label{pif} \

The time variable is discretized as
\begin{equation}
t_{j}=a+j\frac{b-a}{n},~~~j=0,1,2,\cdots ,n.
\end{equation}%
Then, we can show the following with the notations such as $X_j \equiv X(t_j)$ etc.:
\begin{eqnarray}
\int_{a}^{b}dt E[\left\{ D{ X}(t)\right\}  { Y}(t)
+{ X}(t) \tilde{D}{ Y}(t)]
&=& \lim_{n\rightarrow \infty }\sum_{j=0}^{n-1}E\left[ ({X}_{j+1}-%
{X}_{j})\frac{{Y}_{j+1}+{Y}_{j}}{2}\right] \frac{b-a}{n}
\nonumber \\
&&+\lim_{n\rightarrow \infty }\sum_{j=1}^{n}E\left[ \frac{{X}_{j}+%
{X}_{j-1}}{2}({Y}_{j}-{Y}_{j-1})\right] \frac{b-a}{n}
\nonumber \\
&=&\lim_{n\rightarrow \infty }\sum_{j=0}^{n-1}E[{X}_{j+1}{Y}%
_{j+1}-{X}_{j}{Y}_{j}] \frac{b-a}{n} \nonumber \\
&=& \int_{a}^{b}dt E[ \frac{d}{dt} \left\{ X(t)Y(t) \right\}] \nonumber \\
&=&E[{X}(b){Y}(b)-{X}(a){Y}(a)].
\end{eqnarray}
This is called the stochastic generalization of the partial integration formula.

\section{Ito definition or Stratonovich definition?} \label{app:ito-st} \

For usual (non-stochastic) numbers, an integral of a function
$f(x)$ is defined by
\begin{equation}
\int_{x_a}^{x_b} dx f(x) = \sum^{N-1}_{i=0} f(x_i)dx_i ,
\end{equation}
where $dx_i = x_{i+1} - x_i$ and $x_i = x_a +  {\displaystyle  \sum_{j=0}^{i-1} dx_j }$. Here we have used $x_N = x_b$.
However, the right hand side can be re-expressed as
\begin{equation}
\int_{x_a}^{x_b} dx f(x) =\sum^{N-1}_{i=0} f((x_i + x_{i+1})/2)dx_i ,
\end{equation}
because $f((x_i + x_{i+1})/2) = f(x_i + dx_i/2) = f(x_i) + O(dx_i)$ for a general smooth function $f(x)$.
However, these two different definitions give different results when $x$ is a stochastic variable.

Let us denote the Stieltjes integral for the Wiener process as
$\int^t_0 W_s dW_s$.
Then corresponding to the argument above, we can define this integral in two different ways:
one is the Ito definition,
\begin{equation}
(I)\int^t_0 W_s dW_s = \sum_{i=0}^{N-1} W_i (W_{i+1} - W_i),
\end{equation}
and the other is the Stratonovich definition
\begin{equation}
(S)\int^t_0 W_s dW_s = \sum^{N-1}_{i=0} \frac{W_i + W_{i+1}}{2}(W_{i+1} - W_i).
\end{equation}
These two different definitions are known to yield the difference in the results by $t/2$.
Therefore, we must specify the definition of the product for the quantity like $f(W_t) dW_t$.

In the stochastic Lagrangian, we have the term such as $D{\bf r}(t) \cdot D{\bf r}(t)$.
It is very similar to the quantity discussed above, and thus one might insist that we need to
specify one of the definitions for this product. However, $D$ is a conditional expectation
value, and it does not contain any $dW_t$ dependence!
Therefore, we do not need to introduce special definitions for these type of products.
This is also one of the advantages to introduce the mean derivatives.

\section{Ito's lemma (Ito formula)} \label{app:ito} \

Let us assume an arbitrary function $g({\bf x},t)$ which is differentiable for ${\bf x}$ and $t$.
Now we substitute the stochastic variable ${\bf r}(t)$, the solution of Eq.~(\ref{fsde}),
to ${\bf x}$ in $g({\bf x},t)$.
Ito's lemma tells us that this time derivative is given by
\begin{equation}
d g({\bf r}(t),t) = \left( \partial_t  + {\bf u}({\bf r}(t),t)\cdot \nabla + \nu \frac{dt}{|dt|} \nabla^2 \right)g({\bf r}(t),t)dt
+ \sqrt{2\nu}\nabla  g({\bf r}(t),t) d{\bf W}_t .
\end{equation}
Note that this is reduced to the usual Taylor expansion when $\nu = 0$. For stochastic processes, $d{\bf W}_t$ has an order of $\sqrt{|dt|}$, and thus to keep
$O(dt)$, we need to take into account a part of the second order contribution of the Taylor expansion.
When ${\bf r}(t)$ follows Eq. (\ref{bsde}), ${\bf u}$ in the above equation is replaced by $\tilde{\bf u}$.
\\
\\

\end{document}